\documentclass{article}

\usepackage{PRIMEarxiv}

\usepackage[utf8]{inputenc} % allow utf-8 input
\usepackage[T1]{fontenc}    % use 8-bit T1 fonts
\usepackage{hyperref}       % hyperlinks
\usepackage{url}            % simple URL typesetting
\usepackage{booktabs}       % professional-quality tables
\usepackage{amsfonts}       % blackboard math symbols
\usepackage{nicefrac}       % compact symbols for 1/2, etc.
\usepackage{microtype}      % microtypography
\usepackage{lipsum}
\usepackage{fancyhdr}       % header
\usepackage{graphicx}       % graphics
\graphicspath{{media/}}     % organize your images and other figures under media/ folder

\title{Near-Term Enforcement of AI Chip Export Controls Using A Firmware-Based Design for Offline Licensing }

\author{
  James Petrie \thanks{james.petrie@protonmail.com} \\
  \\
}

\begin{document}
\maketitle

\begin{abstract}
Offline Licensing is a mechanism for compute governance that could be used to prevent unregulated training of potentially dangerous frontier AI models. The mechanism works by disabling AI chips unless they have an unused license from a regulator. In this report, we present a design for a minimal version of Offline Licensing that could be delivered via a firmware update. Existing AI chips could potentially support Offline Licensing within a year if they have the following (relatively common) hardware security features: firmware verification, firmware rollback protection, and secure non-volatile memory. Public documentation suggests that NVIDIA’s H100 AI chip already has these security features \cite{vnk_nvidia_2024}. Without additional hardware modifications, the system is susceptible to physical hardware attacks. However, these attacks might require expensive equipment and could be difficult to reliably apply to thousands of AI chips. A firmware-based Offline Licensing design shares the same legal requirements and license approval mechanism as a hardware-based solution. Implementing a firmware-based solution now could accelerate the eventual deployment of a more secure hardware-based solution in the future. For AI chip manufacturers, implementing this security mechanism might allow chips to be sold to customers that would otherwise be prohibited by export restrictions. For governments, it may be important to be able to prevent unsafe or malicious actors from training frontier AI models in the next few years. Based on this initial analysis, firmware-based Offline Licensing could partially solve urgent security and trade problems and is technically feasible for AI chips that have common hardware security features. 
\end{abstract}


\section{Introduction}

Rapid progress on frontier AI systems has raised concerns about the potential for extreme misuse or unintended consequences from AI systems that are trained in the near future. The October 7th export controls \cite{noauthor_implementation_2022} were implemented by the Bureau of Industry and Security (BIS) to regulate access to the AI chips needed to build potentially dangerous AI models. The BIS also requested proposals for technical mechanisms to support these export controls and future regulations \cite{noauthor_implementation_2022}. 
 
Aarne et al. and Kulp et al. proposed a technical mechanism, Offline Licensing, that aims to prevent AI chips from functioning unless they have an up-to-date digital license \cite{aarne_secure_2024, kulp_hardware-enabled_2024}. This mechanism could make AI chip theft or smuggling much more difficult, as stolen or smuggled chips would be rendered useless without a valid license. Moreover, it gives regulators the ability to revoke chip usage if dangerous capabilities are discovered, simply by not re-issuing new licenses.
 
The work by Aarne et al. and Kulp et al. gave a broad overview of technical compute governance mechanisms, and so did not focus on the technical details of a specific Offline Licensing design. In contrast, the aim of this report is to provide a more detailed technical design for the simplest possible Offline Licensing mechanism (while remaining agnostic to which AI chip implements the mechanism). This report is not intended to be authoritative, but rather to present a somewhat opinionated best guess on how to implement Offline Licensing quickly.
 
In Section \ref{sec:TechnicalDesign}, we present the proposed system design, in Section \ref{sec:PotentialAttacks} we discuss potential attacks against the system, and in Section \ref{sec:securityFeatures} we discuss security features needed to make these attacks non-trivial. Finally, in Section \ref{sec:deployment} we present a potential deployment strategy for this technology.
 
In the best case, firmware-based Offline Licensing could be deployed in the next 12 months to new chips and previously sold chips through a firmware update. A more robust hardware-based solution could then be deployed with new chips in 2-4 years \cite{aarne_secure_2024}. The cryptographically signed licenses could be structured identically for both options so that the same authorization process could be used for both. First deploying firmware-based Offline Licensing before switching to hardware based Offline Licensing would safeguard AI chips from misuse much sooner.

\section{Firmware-based Design}
\label{sec:TechnicalDesign}

As described by Aarne et al. and Kulp et al., Offline Licensing is intended to only enable chip usage for chip owners that have regulatory approval \cite{aarne_secure_2024, kulp_hardware-enabled_2024}. To do this, AI chips are modified to receive licenses, which are cryptographically signed messages that specify a compute allowance for each chip. AI chips record the amount of work they have done and halt if it has exceeded the amount allowed by their current license. 
 
There are many ways to implement this mechanism. The design developed here is intended to support current export controls and to be easy to deploy quickly. The specific objectives for our minimal design are to:
\begin{enumerate}
    \item Be unobtrusive to chip owners who have regulatory approval
    \item Make chip usage as difficult as possible for actors without regulatory approval
    \item Be deployable within a year (ideally via a firmware update to already-sold chips)
    \item Make it easy to transition to an improved design in the future (that might have better hardware security or more fine-grained controls) 
\end{enumerate}  
  
Aarne et al. and Kulp et al. identify several metrics that could be used to quantify chip usage. For this design, we will use clock cycles as a usage metric. The reasons to use clock cycles instead of other metrics are: 1) Clock cycles are the closest proxy to chip usage time, which is a simple metric to build regulation around. 2) Our current goal is to either enable or disable chips, not provide fine-grained usage controls. 3) All AI chips can keep track of clock cycles. 4) AI chips likely have several hardware clocks which could potentially be compared for consistency to detect tampering

\begin{figure}[]
\begin{center}
  \includegraphics[width=10cm]{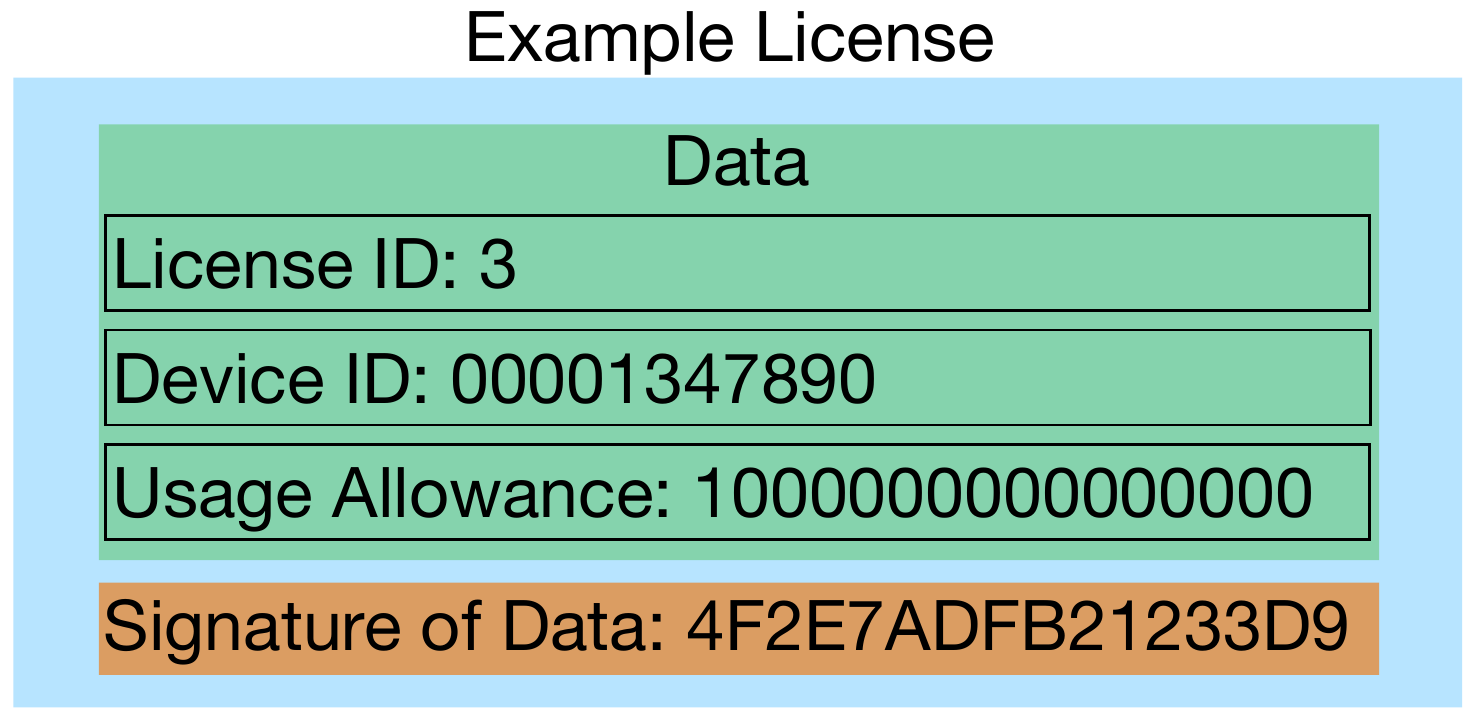}
  \caption{An example license that enables $10^{15}$ clock cycles for a particular chip. The data in this license is cryptographically signed using the private key of the appropriate regulator.}
  \label{fig:exampleLicense}
\end{center}
\end{figure}

The mechanism is built around licenses, an example of which is shown in Figure \ref{fig:exampleLicense}. Each license is a small file that contains: the number of allowed clock cycles, which chip the license is for, and a license ID so that the chip can check that the license hasn’t been used before. Licenses are cryptographically signed using the private key(s) of the appropriate regulator(s) so that the AI chip can verify that the license is authentic. 

\begin{figure}[]
  \centerline{\includegraphics[width=25cm]{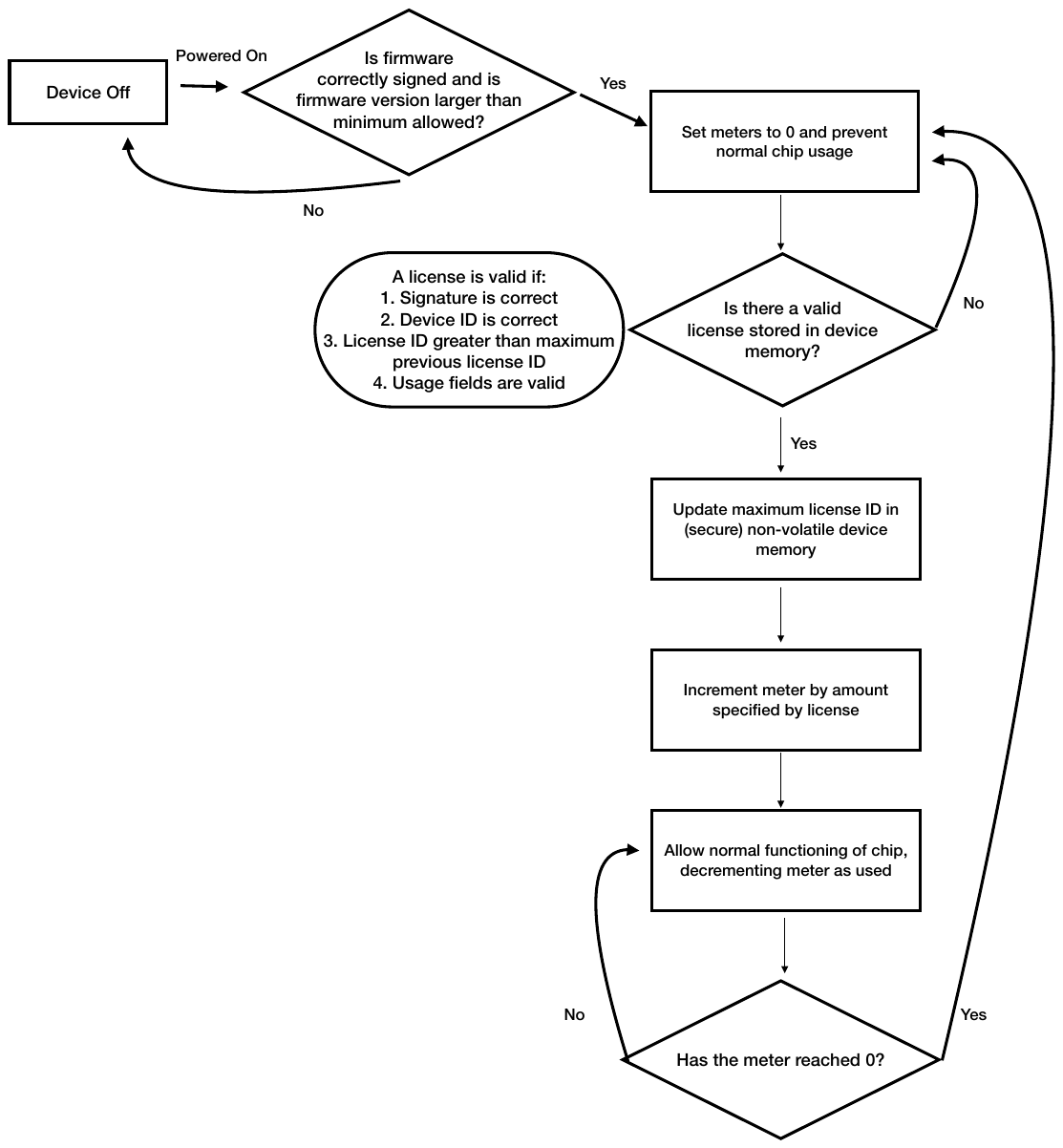}}
  \caption{Control flow for the proposed Offline Licensing design, including firmware verification.}
  \label{fig:controlFlow}
\end{figure}

An outline of the control flow for the mechanism is shown in Figure \ref{fig:controlFlow}. Expanding on the control flow, an AI chip that implements this Offline Licensing mechanism will do the following things (after completing secure boot):

\begin{enumerate}
    \item When an AI chip boots up (or when the allowance on a previous license is exceeded), it checks local non-volatile memory for a valid and unused license  (ideally, multiple licenses could be stored to prevent unnecessarily interrupting chip usage). For the license to be valid and unused: \begin{itemize}
    \item The device ID in the license must match the current device ID
    \item The license ID should be larger than the largest previously used license ID
    \item The usage allowance should be reasonable (e.g., not negative)
    \item The cryptographic signature must correspond to the attached message, and must be signed using a private key that matches the public key of the regulator. The expected public key of the regulator is stored in the firmware binary (and therefore protected from modification by the chip manufacturer’s signature of that binary)
    \end{itemize}

    \item If there isn’t a valid and unused license, the chip sends an error message stating that it has no licenses and requests more. It refuses to complete any computations until a valid license has been provided.

    \item If there is a valid and unused license, it records that that license has been used by securely storing the license ID. The chip then increases the allowed usage meter by the amount specified in the license. The chip can then be used normally.

    \item While the chip is in use, the chip decrements the allowed usage meter as clock cycles are recorded. If the meter reaches 0 the chip halts operation and checks for a new license (return to Step 1).
    
\end{enumerate}

\subsection{Evaluation of Design Objectives}

The control flow in Figure \ref{fig:controlFlow} mostly meets the design objectives, although there are clearly tradeoffs between usability, security, and implementation speed. We expand on these objectives below.

\subsubsection{Unobtrusive to chip owners who have regulatory approval
}
While more obtrusive than having no regulation at all, this design would only require occasional reporting and license entry, most of which could be automated. This would probably be preferable for chip owners compared to a more invasive mechanism that gives regulators remote access for monitoring and control of AI chips. Most AI chips are owned by large companies that can handle these reporting requirements \cite{aarne_secure_2024}.
 
With this implementation, the remaining budget from partially used licenses is lost when an AI chip is powered off \cite{kulp_hardware-enabled_2024}. To retain this budget through a power down without compromising security would require modifiable non-volatile memory that is secure from external writes, as discussed in the Potential Attacks: License Reuse section. 

\subsubsection{Difficult to use AI chips without regulatory approval}

When functioning as intended, this mechanism requires that all chips frequently receive licenses to operate. Stolen chips would not function for long, and smuggled chips would require false license applications to regulators. If a regulator performed a random inspection and discovered chips were missing, they could cease providing licenses for those chips. Additionally, if very dangerous capabilities were detected in a frontier AI system, regulators could delay the allotment of licenses more broadly. 
 
There are several attacks that could potentially be used to evade this system. We discuss these in more detail in Section \ref{sec:PotentialAttacks}, finding that if AI chips have a few common hardware security features then the attacks would likely be difficult to execute.

\subsubsection{Fast deployment timeline}

Overall, this control flow is relatively simple, although the implementation complexity will depend on AI chip architecture details (that are not publicly available). The modifications are probably simpler than for the cryptocurrency mining restrictions \cite{wuebbling_geforce_2021} that NVIDIA implemented in 2021 because there is no need to distinguish between workload types. The needed development time is difficult to accurately estimate without more details about the chips, however a rough estimate is that a firmware-only proof of concept could be developed in 3 months and a more robust version could be developed in 6 months.

\subsubsection{Simple transition to improved system}

The legal framework and license approval process for a design with improved hardware security is nearly identical (as the chip would function the same, just with better protections). Transitioning to a mechanism with more fine-grained usage controls could be done with an additional firmware update.

\section{Potential attacks and countermeasures}
\label{sec:PotentialAttacks}

Here we list potential attacks against this mechanism and design choices that make them more challenging. It is most important to protect against scalable attacks that could be applied at low cost to thousands of chips \cite{kulp_hardware-enabled_2024}. Attacks that require delicate hardware modifications or expensive equipment would be much more difficult to perform for many actors.

\subsection{Firmware modifications or rollback}
  
If attackers could load alternative firmware on an AI chip, then they could circumvent the entire mechanism. A common defense against this is Secure Boot \cite{kelly_hardware_nodate}, which checks that firmware has been signed by the AI chip manufacturer before allowing the chip to boot. With this check, modified firmware would not be able to run.

Another way this attack could be performed is by using older (signed) firmware that doesn’t have the updated licensing mechanism. Fortunately, rollback protection is a common defense against this, which uses a non-decreasing ratchet to keep track of the minimum permissible firmware version . This attack could be prevented by requesting that AI chip owners update the ratchet when first installing the new firmware (and by incrementing it in the factory before selling new chips). It is important to verify afterwards that the ratchet has actually been updated, ideally with device attestation \cite{palmer_attestation_nodate}. 
The idea with that is that the chip collects information about itself and signs it with the device private key (and the chip manufacturer keeps track of each device's public key). If the firmware version ratchet value isn't currently included in the attestation report it could hopefully be added as another field.

Any firmware that is signed and released by the AI chip manufacturer with a version number greater than the value stored in the ratchet can be used. This means that a single firmware release with a vulnerability could be exploited indefinitely by chips that were last updated before that release. To protect against this, each firmware version should be tested thoroughly for vulnerabilities before deployment. If a vulnerability is found in released firmware, the vulnerability should be fixed and chip owners should be instructed to update their firmware (with the ratchet increased so that the old version can no longer be used).

\subsection{License reuse}

Chip owners may attempt to re-use a single license many times, which if successful would provide indefinite chip usage. The simplest defense against this is to record the license ID of licenses that have been used and refuse to accept them in the future. This defense is much less effective if the non-volatile memory where the past licenses are stored can be modified externally (for example, standard flash memory can usually be written to easily).  Cryptographic signing of the license record could prevent trivial modifications, but would not protect against replay attacks (i.e., where the past state is reloaded into the non-volatile memory). 
 
It would be significantly more secure to use the same type of monotonically increasing counter as the firmware version ratcheting to prevent license re-use (especially because it is already a point of failure). This could be achieved by using unused space in the eFuse memory. A potential downside of this approach is that space may be relatively limited (maybe 10s or 100s of unused bits), so this limits the number of licenses that could be allocated in a device lifetime. To make up for this, the license usage amount could be increased so that they don’t need to be reallocated very often.
 
If there is no available and secure non-volatile memory on the chip, then the mechanism could be modified to instead actively request regulator authorization. E.g., at boot-up, each chip could generate a usage request (containing some random data to prevent replay attacks) and the chip owner could send this to the regulator. For pre-authorized chips, this usage request could be immediately signed by the appropriate regulator key and returned to the AI chip. If the authorization is valid, the AI chip could then enable computation for a set number of operations before having to request updated authorization. This approach increases the complexity of the system, but it prevents license reuse attacks by requesting updated authorization instead of trusting a local record of past licenses.

\subsection{Use of same license on multiple chips}
If there is no device ID or the device ID is easy to modify, an attacker could use a single license for thousands of chips. To defend against this, chips should have a unique device ID that is very difficult to modify. 
 
If the chip doesn’t already have a unique device ID, but does have secure non-volatile memory, then a device ID could be randomly generated and stored there. The device ID should be readable by the chip owner so that they can request a license specifically for that chip.

\subsection{Meter Tampering}

Depending on the profiling system used to record compute usage, attackers may be able to prevent compute usage from being recorded accurately by tampering with signal wires (or other physical attacks). This is difficult to prevent in general, but as Kulp et al. note \cite{kulp_hardware-enabled_2024}, collecting multiple readings and performing consistency checks could make this attack much more complicated. For this defense to work, a failed consistency check would have to trigger a destructive action, such as throwing away the remaining compute budget from the current license.

\subsection{Usage Checks Avoided}

The meter readings must be frequently compared with the usage allowance so that the chip knows when to halt. Depending on the design, attackers may be able use the chip in a way that avoids these checks. The risk of this attack could be reduced by tying the usage checks to operations that must occur frequently and are difficult to disable.

\subsection{Halting mechanism disabled}

Depending on how the halting mechanism is implemented, it may be possible for attackers to disable it when the compute budget has been used. Defenses against this will depend on the implementation. 

\subsection{Regulator Private Key Leaked}

If chip owners can acquire the private key of the regulator, they can generate licenses that appear to be authentic. For this reason, it is very important to keep the regulator’s private key secure. A potential way to mitigate this risk would be to require multiple different regulators to sign each license, or potentially a fraction of approved regulators (e.g. $\frac{2}{3}$). This is not just a technical decision, but also depends on which organizations are responsible for regulating AI chip usage. 
 
Similarly, the private key used by the chip manufacturer to sign firmware must also be kept secure, because actors that have this private key could make modified firmware appear legitimate. This attack could either be performed by stealing the private key, or by getting an employee at an AI chip manufacturer to introduce a backdoor or create a modified firmware version.

%\subsection{Regulator Private Key Deleted}

%todo

\subsection{License Stockpiling}

Chip owners could save licenses instead of using them, which would allow them to use AI chips for a longer time in a situation where license approvals are halted. To make this more difficult to do, device attestation showing that licenses have been used could be required by regulators before allocating more licenses. 

%\subsection{Introduction of Backdoors}

%already at risk of this - mechanism doesn't change

\section{AI Chip Security Features}
\label{sec:securityFeatures}

Based on the attack scenarios described above, the hardware security features marked as high importance in Table \ref{table:security} are needed to make attacks nontrivial. Additional security features that offer further protection are also included.

\begin{table}
 \caption{Hardware Security Features}
  \centering
  \begin{tabular}{p{4cm}|p{7cm}|p{4cm}}
    \toprule
    Hardware Security Feature   & Reason for Security Feature     & Importance \\
    \midrule
    \midrule
    Firmware verification and rollback protection & Prevent chip usage with modified firmware or earlier versions of firmware that don’t have restrictions  & High     \\
    \midrule
    Non-volatile memory that is difficult for an attacker to modify     & To record the use of licenses so that they cannot be used again. \newline Also, to store a unique device ID so that licenses can be allocated to a specific chip.  & High (or instead require the chip to request a new license each time it boots up)     \\
    \midrule
    Device attestation of the firmware rollback version and license usage     & Provide evidence to regulators that the chip can no longer use insecure firmware and that licenses are being used and not stockpiled       & Medium  \\
    \midrule
    Redundant meters measuring chip usage     & Make physical attacks more difficult by performing a consistency check on meter readings     & Medium \\
    \midrule
    Tamper evidence / tamper proofing     & Detect physical attacks during random inspections, or automatically destroy the chip if a physical attack is attempted     & Medium \\
    \bottomrule
  \end{tabular}
  \label{table:security}
\end{table}

\subsection{Security features on current AI chips}

Public documentation on security features for AI chips is quite limited, however NVIDIA’s H100 Product Brief says that \cite{vnk_nvidia_2024}:
\begin{quotation}
\noindent
“The NVIDIA H100 NVL provides secure boot capability using CEC. Implementing code authentication, rollback protection and key revocation, the CEC device authenticates the contents of the GPU firmware ROM before permitting the GPU to boot from its ROM.”
\end{quotation}

Rollback protection requires secure and non-volatile memory to store the minimum firmware version, so this documentation implies that the H100 has all of the “high” importance security features needed to implement Offline Licensing. However, it isn’t clear how much remaining space is available within secure memory that could be used for recording used licenses.

\section{Deployment Feasibility}
\label{sec:deployment}

The mechanism described in this report could aid the US in safeguarding dangerous technology and enforcing the October 7th export controls. NVIDIA and other AI chip companies may be incentivized to implement this mechanism if it allowed them to sell chips to customers that would otherwise be prohibited by export controls (either the current export controls for China or other more expansive export controls in the future).
 
The design should ideally be reviewed by auditors and red-teamed by researchers that are not affiliated with AI chip manufacturers. This is because AI chip manufacturers may have differing priorities (i.e., to develop the mechanism at low cost and to minimize disruptions to their customers).
 
Once the Offline Licensing firmware is ready, new chips could be automatically loaded with this firmware and rollback protection could be set to prevent the use of old firmware. For chips that have already been sold, government regulation compelling a firmware update could be effective if possible. If this is not possible, chip owners could be incentivized to comply by making this a requirement in order to retain their BIS “presumption of approval” status for future chip purchases. Updating the firmware may also be required in order to retain a device warranty.  

Chip owners that do update their firmware could demonstrate this with device attestation \cite{palmer_attestation_nodate}. This attestation would ideally include evidence that the firmware ratchet has been increased so that previous firmware versions cannot be used.
 
The regulating organization(s) who make license decisions are left unspecified here, but it should probably not be AI chip manufacturers because of their potential conflict of interest (e.g. avoiding bad PR from denying licenses). However, AI chip manufacturers could potentially be responsible for distributing approved licenses from their existing servers (to minimize the technical burden on the regulator). 
 
The information needed to decide on license approvals could be sent directly from chip owners to the regulator.  Chip owners could request licenses for air-gapped chips by collecting the device IDs of their chips and sending them to the regulator. At the very simplest this could be done by email, or a dedicated web portal. The license application could optionally require information about the chip owner, usage purpose, location, etc. 
 
If the regulator verifies that a chip owner is authorized to be using the chips, then they can generate licenses for them using their private key. The number of licenses per chip can be chosen to reduce the regulatory burden on chip owners while retaining the ability to halt AI chip usage in an emergency. These licenses could then be sent to AI chip manufacturers to be distributed to chip owners. This structure would prevent AI chip manufacturers from accidentally approving licenses, but would not place much of the burden of maintaining technical infrastructure on the regulator.
 
If possible, unannounced inspections of AI chips could be performed to check for evidence of tampering and to verify that chips are where they are supposed to be. Discovering unexpected tampering might indicate that a chip owner had attempted to physically modify a chip to allow unregulated usage. Tamper-evident packaging could potentially be added to chips sold in the future to aid this type of inspection. 
 
A bug-bounty program could be set up to reward researchers for discovering and privately disclosing flaws in the Offline Licensing system. If legally possible, having AI chip manufacturers pay for the bounties could serve as an incentive for them to ensure the design is secure. Alternatively, making the implementation secure could be incentivized by relaxing export controls on these chips if significant bugs are not found. The bug bounty program could potentially be run by DARPA as with their \$20M AI cybersecurity challenge \cite{gill_darpa_2023}.

\section{Conclusion}

For AI chips with common security features, it could be possible to implement Offline Licensing with only a firmware update within a year. While a firmware-only solution is not fully secure, it could be robust enough to make workarounds unreliable or very expensive. Quickly deploying Offline Licensing would make AI chip theft and smuggling much more difficult and could allow regulators to revoke chip access in an emergency.

\section*{Acknowledgments}

Thanks to Onni Arne and Gabriel Kulp for advice and thorough reports on Offline Licensing, Christopher Phenicie for helpful comments and information about the BIS, Konstantin Pilz for his suggestion to look for mechanisms that only require firmware modifications, and Lennart Heim and Asher Brass for useful feedback on this report.

%Bibliography
\bibliographystyle{unsrt}  
\bibliography{references}

\end{document}